# Quantify Influence of Delay in Opinion Transmission of Opinion Leaders on COVID-19 Information Propagation in the Chinese Sina-microblog

Fulian Yin[1], Xueying Shao[1], Meiqi Ji[1], Jianhong Wu[2,*]

[1] College of Information and Communication Engineering, Communication University of China, Beijing, 100024, PR China
[2] Fields-CQAM Laboratory of Mathematics for Public Health, Laboratory for Industrial and Applied Mathematics, York University, Toronto, M3J1P3, Canada
[*] Correspondence: Email: wujhhida@hotmail.com; Tel: +1-416-736-5243; Fax: +1-416-736-5698.

**Abstract:**

In a fast evolving major public health crisis such as the COVID-19 pandemic, multiple pieces of relevant information can be posted sequentially in a social media platform. The interval between subsequent posting times may have different impact on the transmission and cross-propagation of the old and new information to result in different peak value and final size of forwarding users of the new information, depending on the content correlation and whether the new information is posted during the outbreak or quasi steady state phase of the old information. To help in designing effective communication strategies to ensure information is delivered to the maximal number of users, we develop and analyze two classes of susceptible-forwarding-immune information propagation models with delay in transmission, to describe the cross-propagation process of relevant information. We parametrize these models using real data from the Sina-Microblog and use the parametrized models to define and evaluate mutual attractiveness indices, and we use these indices and parameter sensitivity analyses to inform strategies to ensure optimal strategies for a new information to be effectively propagated in the microblog.

*Keywords:* COVID-19, delay transmission, dynamic model, Sina-Microblog

## 1. Introduction

In the absence of effective treatment and vaccine, the success or failure of mitigating COVID-19 transmission in the population relies heavily on the effectiveness of social distancing, self-protection, case detection, quarantine, isolation and testing. The effectiveness of these non-pharmaceutical interventions depends on the active participation and engagement of residents in the community, which is significantly influenced by the public opinions. Given the very short disease transmission doubling time, the timing (and hence time lags) in the communication of critical public health information to the community has a profound impact on the outcome of public adherence to non-pharmaceutical measures, and ultimately on the outcome of outbreak mitigation. Adding to the challenging of effective communications is the cross-propagation of relevant, and sometimes inconsistent, pieces of information that enter social media at different time points. This calls for a strategy of optimizing the timing of posting key information in social media during a fast evolving pandemic.

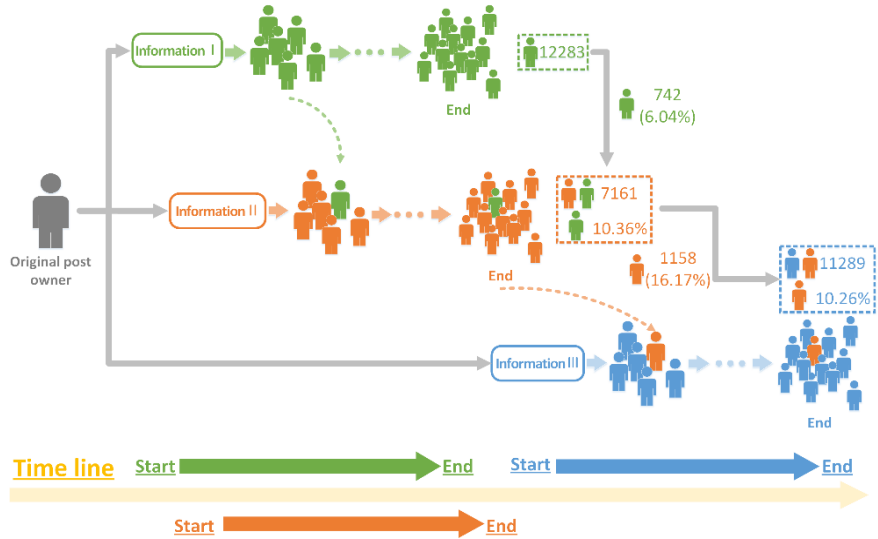

**Figure 1.** Cross-propagation of three relevant pieces of information related to COVID-19 pandemic, posted in sequence on Sina-microblog.

Figure 1 shows the cross-propagation of three pieces of related information about COVID-19: titled by "Just want a regular 20200202", "Announcement of donation acceptance by Wuhan JinYinTan hospital" and "Three cases of community transmission were reported in Shenzhen for the first time". These pieces of information were posted in Sina-microblog, with different beginning and ending time points marked in the (horizontal) timeline. Almost immediately after reading Information A, some users forwarded the Information B, so both pieces of information shared similar life cycles with respective beginning and ending time points close to each other. There were 12283 Information A users, among which 742 (6.04%.) forwarded Information B. Eventually, 7161 of users forwarded Information B, and those who simultaneously forwarded Information A accounted for 10.36%. Information C was then released and users of Information B started to forward Information C. At the end, among 7161 users of Information B, 1158 (16.17%) also forwarded Information C, accounting for 10.26% of the users for C.

In general, relevant information, when posted with an appropriate time lag, can attract the interest of social media users in public hot events by increasing the efficiency of dissemination of a group of information cross-propagated. It is an important topic of research, the main objective of our study, to understand the information cross-propagation dynamics in order to inform optimal strategies of posting a relevant information in an appropriate time sequence to ensure their maximum interaction for effective co-promotion during a public health emergency situation.

To our best knowledge, no appropriate model framework has been developed and analyzed to examine the impact of information cross-propagation dynamics for a group of relevant information posted subsequently. Here, we try to fill this gap by proposing a susceptible-forwarding-immune model with time-delayed posting and transmission. We develop this framework, illustrate this framework and parametrize our model by using the forwarding quantity that represents public attention to some popular opinions on the COVID-19 pandemic. Our focus is on the dynamic interactions among several pieces of information posted sequentially, and we aim to examine the impact of time lags between different posting time points on the evolution and steady states of cross-propagation.

## 2. Related works

In the field of information propagation dynamics, there is a strong similarity between the propagation of rumors and pathogens spread in the population [1]. Much has been done that uses epidemic models to examine rumor propagation in the hope that the negative influence of rumors can be eliminated or at least minimized. For example, susceptible-infected-exposed-recovered model [2-3], susceptible-infected model [4-5], susceptible-infected-susceptible (SIS) [6] model and susceptible-infected-recovered model [7-8], have all been developed and recognized as among classical propagation dynamics models.

The development of the Internet and the enrichment of social media mandate further extensions of traditional models to reflect novel transmission mechanisms and to take advantage of data from multiplatform. Gu and Cai [9] and Gu et al. [10] proposed the forget-remember mechanism to study the spreading process in a 2-state model. Zhao et al. [11] combined the forgetting mechanism and the SIR model to represent the rumor spreading process in an online social blogging platform LiveJournal. In 2014, Zhao et al. [12] integrated the refutation mechanism in homogeneous social networks into the SIR model and analyzed the dynamic process of rumor propagation. Considering the three influencing factors of enterprises affected by rumors, pinion leaders and microblog platform, an SIR model based on browsing behavior was constructed to explain how rumors spread among followers under the influence of different rumor refuting measures [13]. Other features of the new media were further incorporated in Zhao et al [14]. Borge-Holthoefer et al. [15] considered the case when spreaders were not always active and an ignorant was not interested in spreading the rumor, and then separately introduced these ideas into two different models. They concluded that these models provided higher adhesion to real data than classical rumor spreading models. In 2020, Yin et al [16] considered the user's behavior of re-entering new topics and proposed a multiple-information susceptible-discussing-immune (M-SDI) model to investigate COVID-19 relevant information propagation in the Chinese Sina-microblog. Ding et al. [17] proposed an improved SIR model, which used differential equations to study the rule of information transmission on media platform and predict microblog information accurately. Wang et al. [18] proposed a modeling method that considers Weibo propagation behavior based on SIS model, the forwarding trend in the future can be predicted. Zhang et al. [19] focused on the impact of media transmission and interpersonal relationships on information propagation and then proposed the media and interpersonal relationship-SEIR (MI-SEIR) model. Zhao et al. [20] developed a new rumor spreading model called susceptible-infected-hibernator-removed (SIHR) model, introducing a new kind of people-hibernators in order to reduce the maximum rumor influence. Woo et al. [21] proposed an event-driven SIR model based on the impact of news release on social media to reflect the impact of specific events on opinion diffusion. Yao et al. [22] concentrated on examining the influence of different interactions among information on the spread of public opinion and modeling based on SIR model, which verified the otherness of public opinion under a distinct information environment. Other studies relevant to our work include [23-26]. In particular, Tanaka M et al. [26] added a new module to the traditional model by using two datasets from the Japanese Mixi and Facebook rather than a single dataset.

Many researches on cross-transmission in disease diffusion are highly significant to study the cross-propagation of information. Feng et al. [27] established a mathematical model that incorporated the virus mutation dynamics in the transmission of CHIKV among mosquitoes and humans. However, the important phenomenon of time lag in posting and cross-propagation of relevant information for information dissemination in real social media networks has not been adequately addressed in these earlier studies. We noted that Zan et al. [28] studied the double rumors spreading with different launch times, in which the new rumor was launched with a certain delay but also could interact with the old rumor. Zan et al proposed two classes of double-rumors spreading models: a double-susceptible-infected-recovered (DSIR) model, where it was assumed that the rumor was disseminated by direct contacts of infective nodes with others; and a comprehensive-DSIR (C-DSIR) model, with which the authors studied the whole spread situation of all rumors with a focus on determining how many people did not spread all rumors in the entire period or how many were spreading or had spread at least one of rumors.

In comparison with the aforementioned studies, here we consider the phenomenon where at different propagation stages of a piece of information posted in the social media, other pieces of information are posted and their relevance in contexts and posting time sequence combined generate outbreak for each piece of information, and more importantly cross-propagation in which users of one piece of information forward other pieces of information they are exposed later. We develop two classes of dynamic propagation models that focus respectively on the single information transmission and multi-information cross-propagation patterns during explosive and quasi steady state periods of the information posted sequentially. We aim to examine emphatically the influence of different participating groups of a posted information on the spread of the information from the participating groups. As populations who have forwarded, exposed/or not exposed to relevant posted information are attracted to a new piece of information differently, by introducing and analyzing the impact of *attractiveness indexes* on relevant information propagation, and examining the significant factors of delaying in posting relevant information propagation, we inform strategies for sequentially posting relevant information to achieve effective communication of key public opinions. We will illustrate this with data from public opinions about COVID-19 pandemic management.

3.  Case study on COVID-19 opinion leader information sequential release in Sina-microblog

Since the COVID-19 outbreak in China, intensive information which were clearly relevant to each other has been frequently posted. Figure 2 shows the total forwarding quantities of their Weibos during each one-hour time frame on January 25, 2020, and January 26, 2020 (Data-acquisition up to February 19, 2020), for top ten opinion leaders of this outbreak event in the Chinese Sina-microblog. As illustrated, those pieces of information with high influence were posted extremely frequently by each opinion leader. At the same time, there was a strong correlation among a series of Weibos posted by these opinion leaders during certain periods. Of all the data showed in Figure 2, People's Daily issued five Weibos in one hour from 22:00 to 23:00 on January 25, and they were forwarded by a total of more than 300,000 users. Therefore, the frequent release of relevant information by original post owners was a common phenomenon in the COVID-19 information propagation, and understanding its effectiveness is important.

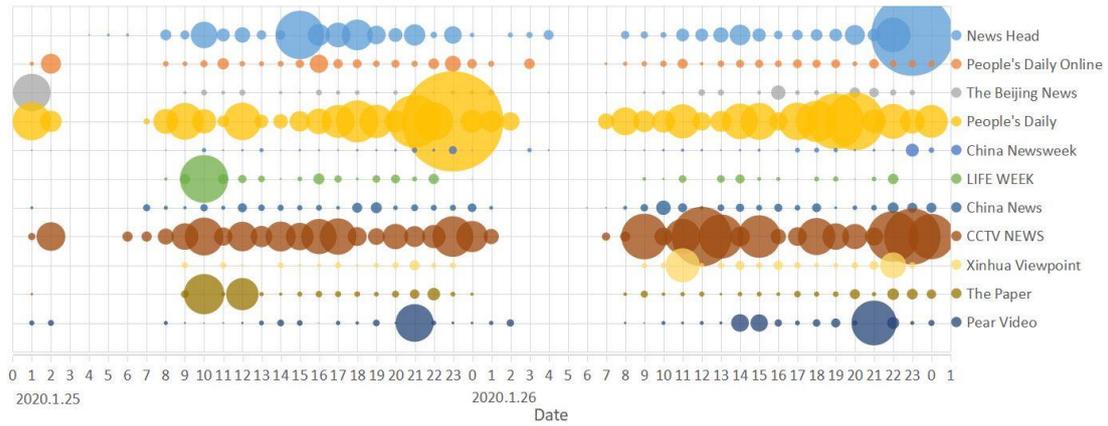

**Figure 2.** A bubble chart of forwarding quantity of public opinions on COVID-19 information by top ten opinion leaders in Weibos, during an early period of COVID-19 outbreak in China.

**Table 1**. Cumulative forwarding quantity of Information A posted at 8:41 on February 2, 2020

| Time | 0 | 10min | 20min | 30min | 40min | 50min | 60min | 70min |
|---|---|---|---|---|---|---|---|---|
| Information A | 47 | 597 | 940 | 1208 | 1458 | 1691 | 1937 | 2182 |
| Time | 80min | 90min | 100min | 110min | 2h | 3h | 4h | 5h |
| Information A | 2477 | 2952 | 3461 | 3917 | 4390 | 6366 | 7501 | 8281 |
| Time | 6h | 7h | 8h | 9h | 10h | 11h | 12h | 13h |
| Information A | 8846 | 9293 | 9638 | 9954 | 10199 | 10435 | 10795 | 11138 |
| Time | 14h | 15h | 16h | 17h | 18h | 19h | 20h | 21h |
| Information A | 11459 | 11812 | 12013 | 12088 | 12109 | 12119 | 12128 | 12136 |
| Time | 22h | 23h | 24h | 25h | 26h | | | |
| Information A | 12140 | 12146 | 12157 | 12171 | 12184 | | | |

**Table 2**. Cumulative forwarding quantity of Information B posted at 10:41 on February 2, 2020

| Time | Around 2h | 3h | 4h | 5h | 6h | 7h | 8h |
|---|---|---|---|---|---|---|---|
| Information B | 15 | 1281 | 2615 | 4013 | 4817 | 5322 | 5685 |
| Time | 9h | 10h | 11h | 12h | 13h | 14h | 15h |
| Information B | 5932 | 6052 | 6152 | 6264 | 6317 | 6380 | 6401 |
| Time | 9h | 17h | 18h | 19h | 20h | 21h | 22h |
| Information B | 6423 | 6434 | 6447 | 6454 | 6455 | 6456 | 6458 |
| Time | 23h | 24h | 25h | 26h | | | |
| Information B | 6460 | 6461 | 6465 | 6471 | | | |

**Table 3**. Cumulative forwarding quantity of information C posted at 17:51 on February 2, 2020

| Time | Around 9h | 10h | 11h | 12h | 13h | 14h | 15h |
|---|---|---|---|---|---|---|---|
| Information C | 20 | 1180 | 4244 | 6235 | 7595 | 8572 | 9103 |
| Time | 16h | 17h | 18h | 19h | 20h | 21h | 22h |
| Information C | 9381 | 9569 | 9642 | 9680 | 9700 | 9736 | 9764 |
| Time | 23h | 24h | 25h | 26h | | | |
| Information C | 9800 | 9864 | 9943 | 10006 | | | |

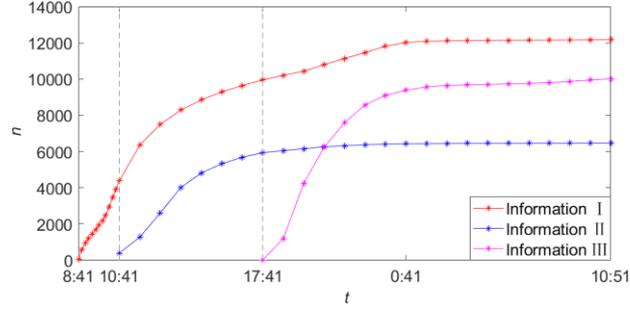

**Figure 3.** Cumulative forwarding quantity of three pieces of information.

Figure 3 shows the trend of cumulative forwarding users for the three pieces of information in Table 1-3. It shows that when Information A broke out, Information B was posted almost immediately. Compared with the Information, the outbreak period of information B was shorter and the trend was flatter. Information C was released during the quasi steady-state period of information B. In comparison with Information, the outbreak period of Information C lasted longer, meanwhile, the cumulative forwarding quantity was also larger.

Sequentially releasing two related pieces of information by the same original post owners within the same COVID-19 theme was a common phenomenon. Also importantly, different entering times of new information during the spreading process of an old (previously posted) information exhibited different promoting effects on the cross-propagation and cross-promotion of relevant public opinions. Here we focus on users who have exposed one posted information may have a special interest in, and hence susceptible to, new and relevant information. This represents a remarkable difference from the spread of rumors and other traditional public hot events. Our information cross-propagation delay in transmission model (DT-SFI). including short interval delay in transmission susceptible-forwarding-immune (STI DT-SFI) dynamics model and large interval delay in transmission susceptible-forwarding-immune (LTI DT-SFI) dynamics model, is developed to take into consideration the situations when the relevant information is posted during the outbreak period, or during the quasi steady-state period of the previously posted (old) information.

## 4. Large delay in transmission susceptible-forwarding-immune dynamics model

### 4.1 Model description

The structure of our proposed large interval delay in transmission susceptible-forwarding-immune (LTI DT-SFI) dynamics model is shown in Figure 4. There are two phases involved: In phase 1, a piece of stand-alone information (Information 1) is spreading; and in phase 2, another piece of information (Information 2) is posted during the quasi steady-state period of the posted information.

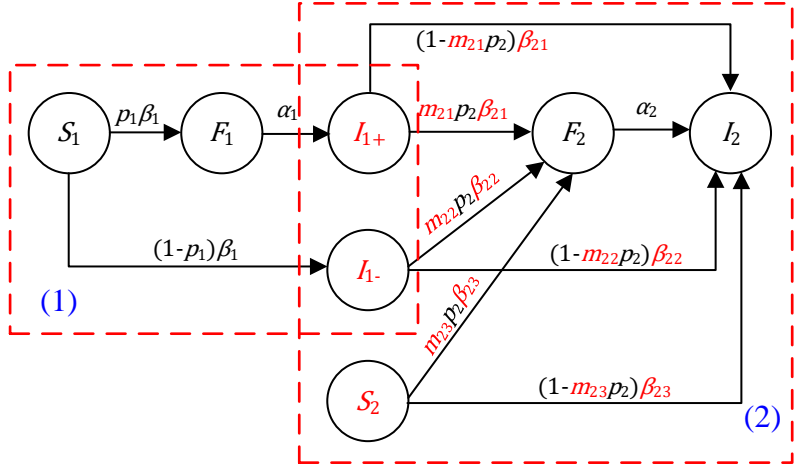

**Figure 4**. A schematic illustration of the information cross-spreading, where a new piece of information is posted during the quasi steady-state period of an already posted information.

**Phase 1. Post a stand-alone information**: The propagation dynamics during phase 1 for one posted information is modeled based on the traditional SFI [29] model, with a novel stratification of the immune population. Namely, there will be two classes of immune populations (as far as Information 1 is concerned): those who have forwarded the posted information but are no longer in their active period forwarding this posted information ($I_{1+}$), and those who have exposed to the information but are not interested in forwarding this ($I_{1-}$). This distinction of immunity is important as individuals in these two distinct compartments will have different levels of interest in other relevant information that will be posted later. This will allow us to introduce different measures of attractiveness to new relevant information.

So, in our model, we stratify the population ($N_1$) into four states: the susceptible state of the posted information ($S_1$), the forwarding state of the posted information ($F_1$), the inactive immune state ($I_{1+}$), the direct immune state ($I_{1-}$). A susceptible user can be exposed to the posted information with an average exposure rate $\beta_1$ and will forward the information with the forwarding probability $p_1$. The forwarding users can become inactive immune users with an average rate $\alpha_1$. So, a user may have a unique state, with $S_1(t)$, $F_1(t)$, $I_{1+}(t)$ and $I_{1-}(t)$ denoting the number of users in the susceptible, forwarding, inactive and direct immune state, respectively. We obtain the following DT-SFI dynamics model in phase 1:

$$\begin{cases} dS_1(t)/dt = -\beta_1 S_1 F_1 \\ dF_1(t)/dt = p_1 \beta_1 S_1 F_1 - \alpha_1 F_1 \\ dI_{1+}(t)/dt = \alpha_1 F_1 \\ dI_{1-}(t)/dt = (1 - p_1)\beta_1 S_1 F_1 \end{cases} \quad (1)$$

The state transition of different populations in phase 1 can be interpreted as follows: An active forwarding user will contact an average number of $\beta_1 N_1$ users per unit time and the probability of a user is a susceptible user of the posted information is $S_1(t)/N_1$, so an active forwarding user will contact $\beta_1 S_1(t)$ susceptible users, among which $p_1 \beta_1 S_1(t) F_1(t)$ susceptible users will choose to forward the information and $(1 - p_1)\beta_1 S_1(t) F_1(t)$ will not. As time goes by, $\alpha_1 F_1(t)$ will go to the immune state from the forwarding period when they do not influence other users as far as Information 1 is concerned.

For the data fitting purpose, we note that the Sina-microblog provides an important data about any piece of information relevant to COVID-19, the number of cumulative forwarding quantity, given by

$$C_1(t) = \int_0^t p_1 \beta_1 S_1 F_1 \, dt. \tag{2}$$

**Phase 2. Post a new information during the quasi steady-state period of the posted information:** Now we consider that a piece of new information is posted at time $t_\tau$ when the posted information is already in the quasi steady-state. We introduce

- *an extensive exposure attractiveness index*, for the individuals in the immune state who have forwarded the posted information but are no longer in their active forwarding period. These individuals are more susceptible to the new yet relevant information as they had interest in the posted information;
- *a mild exposure attractiveness index*, for the direct immune individuals. These individuals have had exposure to the posted information but had shown little interest to information;
- *an un-exposure attractiveness index*, for those who have never exposed to the first posted information.

Accordingly, we introduce three states of the population ($N_2$) for the newly posted information: the susceptible state ($S_2$), the forwarding state ($F_2$), and the immune state ($I_2$). We summarize the notations in Table 4.

**Table 4**. Parameters definition

| Parameter | Interpretation |
|---|---|
| **Attractiveness parameters, stratified by the exposure to old information** | |
| $m_{21}$ | The extensive exposure attractiveness index that an inactive user of state $I_{1+}$ becomes a forwarding user of state $F_2$. |
| $m_{22}$ | The mild exposure attractiveness index that a direct immune user of state $I_{1-}$ becomes a forwarding user of state $F_2$. |
| $m_{23}$ | The un-exposure attractiveness index that a new susceptible user of state $S_2$ becomes a forwarding user of state $F_2$. |
| **Transmission parameters associated with the different attractiveness** | |
| $\beta_{21}$ | The average exposure rate that the inactive users of the old information can contact with the newly posted information. |
| $\beta_{22}$ | The average exposure rate that the direct immune users of the old information can contact with the newly posted information. |
| $\beta_{23}$ | The average exposure rate that the new susceptible users can contact with the newly posted information. |
| $p_2$ | The probability that an exposed user will forward the newly posted information. |
| $\alpha_2$ | The average rate at which a user in the forwarding state of newly posted information becomes inactive to forwarding, where $1/\alpha_2$ is the average duration a forwarding user remains active in forwarding newly posted information. |

Each user may have a unique state, with $I_{1+}(t)$, $I_{1-}(t)$, $S_2(t)$, $F_2(t)$ and $I_2(t)$ denoting the number of users in the susceptible, forwarding and immune state, respectively. We obtain the following LTI DT-SFI dynamics model in phase 2:

$$\begin{cases} dS_2(t)/dt = -\beta_{23}S_2F_2 \\ dI_{1+}(t)/dt = -\beta_{21}I_{1+}F_2 \\ dI_{1-}(t)/dt = -\beta_{22}I_{1-}F_2 \\ dF_2(t)/dt = m_{21}p_2\beta_{21}I_{1+}F_2 + m_{22}p_2\beta_{22}I_{1-}F_2 + m_{23}p_2\beta_{23}S_2F_2 - \alpha_2 F_2 \\ dI_2(t)/dt = (1-m_{21}p_2)\beta_{21}I_{1+}F_2 + (1-m_{22}p_2)\beta_{22}I_{1=}F_2 + (1-m_{23}p_2)\beta_{23}S_2F_2 + \alpha_2 F_2 \end{cases}$$

(3)

The mass action in phase 2 can be interpreted as follows: An active forwarding user will contact an average number of $\beta_{21}N_2$ inactive immune users of posted information per unit time and the probability of a user is an inactive immune user is $I_{1+}(t)/N_2$, so an active forwarding user will contact $\beta_{21}I_{1+}(t)$ inactive immune users, among which $m_{21}p_2\beta_{21}I_{1+}(t)F_2(t)$ will choose to forward the new information and $(1-m_{21}p_2)\beta_{21}I_{1+}(t)F_2(t)$ will not; an active forwarding user will contact an average number of $\beta_{22}N_2$ direct immune users of posted information per unit time and the probability of a user is a direct immune user is $I_{1-}(t)/N_2$, so an active forwarding user will contact $\beta_{22}I_{1-}(t)$ direct immune users, among which $m_{22}p_2\beta_{22}I_{1-}(t)F_2(t)$ will choose to forward the new information and $(1-m_{22}p_2)\beta_{22}I_{1-}(t)F_2(t)$ will not; an active forwarding user will contact an average number of $\beta_{23}N_2$ susceptible users of the newly posted information per unit time and the probability of a user is a susceptible user is $S_2(t)/N_2$, so an active forwarding users will contact $\beta_{23}S_2(t)$ susceptible users, among which $m_{23}p_2\beta_{23}S_2(t)F_2(t)$ will choose to forward the new information and $(1-m_{23}p_2)\beta_{23}S_2(t)F_2(t)$ will not.

The forwarding quantity of the newly posted information is given by

$$C_2(t) = \int_0^t (m_{21}p_2\beta_{21}I_{1+}F_2 + m_{22}p_2\beta_{22}I_{1-}F_2 + m_{23}p_2\beta_{23}S_2F_2)\,dt. \tag{4}$$

**The public opinion reproduction ratio $\mathfrak{R}_o$**: Since the newly posted Weibo starts at different times and develops differently under the influence of prior information, we define the information reproduction ratio

$$\mathfrak{R}_o = \frac{m_{21}p_2\beta_{21}I_{1+,\tau} + m_{22}p_2\beta_{22}I_{1-,\tau} + m_{23}p_2\beta_{23}S_{20}}{\alpha_2}. \tag{5}$$

This is the total number of Information 2 users generated by introducing a typical Information 2 user, at time $\tau$ after Information 1 was posted, during its entire period of active forwarding. The initial population for Information 2 has been stratified by the exposure of the entire population to Information 1 during the time interval $[0, \tau]$. The relative size of $\mathfrak{R}_o$ to the unity determines if Information 2 can generate an information outbreak: since $F_2'(0) = (m_{21}p_2\beta_{21}I_{1+,\tau} + m_{22}p_2\beta_{22}I_{1-,\tau} + m_{23}p_2\beta_{23}S_{20} - \alpha_2)F_2(0)$. We conclude that $F_2'(0) > 0$ if and only if $\mathfrak{R}_o > 1$.

### 4.2 Data fitting

**Parameter estimation**:

To fit our model with real data from the Sino-microblog, we use the LS method to estimate the LTI DT-SFI model parameters and the initial data. In phase 1, the parameter vector is set as $\Theta_1 =$ $(p_1, \beta_1, \alpha_1, S_{10})$, and the corresponding numerical calculation based on the parameter vector for $C_1(t)$ is denoted by $f_{C_1}(k, \Theta_1)$. The LS error function

$$LS = \sum_{k=0}^{T} |f_{C_1}(k, \Theta_1) - C_{1k}|^2 \tag{6}$$

is used in our calculation, where $C_{1k}$ denotes the actual cumulative forwarding populations of the posted information. Similarly, in phase 2, the vector is set as

$\Theta_2 = (\beta_{21}, \beta_{22}, \beta_{23}, m_{21}, m_{22}, m_{23}, p_2, \alpha_2, S_{20})$,

and the corresponding numerical calculation based on the parameter vector for $C_2(t)$ is denoted by $f_{C_2}(k, \Theta_2)$. The LS error function

$$LS = \sum_{k=0}^{T} |f_{C_2}(k, \Theta_2) - C_{2k}|^2 \tag{7}$$

is used in our calculation, where $C_{2k}$ denotes the actual cumulative forwarding populations of the newly posted information. Here, $n = 1,2 ...$ represents the different phases, $k = 0,1,2,...$ is the sampling time, $n = 1,2,3$. We estimate the parameters of our **LTI DT-SFI** model with the data of Information B and Information C.

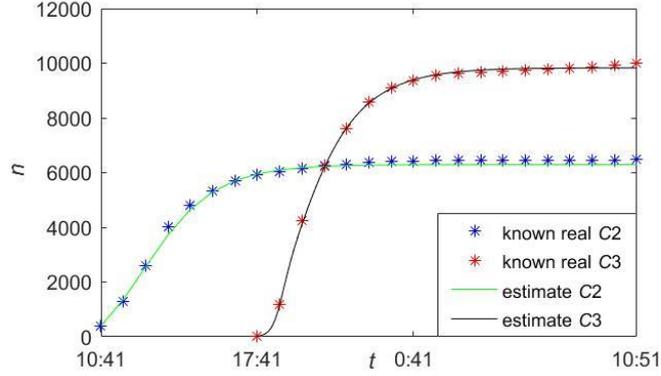

**Figure 5.** The data fitting results of Information B and Information C.

Figure 5 reports our data fitting results for Information B and Information C on the real data given in Table 2-3, where the blue star denotes the actual cumulative number of forwarding users of Information B, the red star denotes the actual cumulative number of forwarding users of Information C, the green line and the black line denotes the estimated cumulative number of forwarding users of Information B and Information C, respectively.

**Table 5**. Values of some important parameters, estimated for Information B

| Name | $S_{10}$ | $\alpha_1$ | $\beta_1$ | $p_1$ |
|---|---|---|---|---|
| Value | $5.6458 \times 10^6$ | 1.5757 | $1.7901 \times 10^{-4}$ | 0.0020 |

**Table 6**. Values of important parameters, estimated for Information C

| name | $\beta_{21}$ | $\beta_{22}$ | $\beta_{23}$ | $m_{21}$ | $m_{22}$ | $m_{23}$ | $p_2$ | $\alpha_2$ | $S_{20}$ |
|---|---|---|---|---|---|---|---|---|---|
| value | 0.8994 | 0.0023 | $1.0871 \times 10^{-4}$ | 0.2468 | 1.9895 | 0.5559 | $2.9516 \times 10^{-4}$ | 0.9858 | $7.4439 \times 10^6$ |

Table 5 and Table 6 give estimated values of important parameters for Information B and Information C, respectively. We can see in phase 2, when Information C was posted during the quasi steady-state period of Information B, the average exposure rate $\beta_{21}$ was the largest, indicating that an inactive user of the posted Information B was more susceptible to the newly posted Information C; the average exposure rate $\beta_{23}$ was small, indicating that a susceptible user of the newly posted Information C contacted the information at a lower rate. In addition, among the three attractiveness indices, the index $m_{22}$ is the largest, which indicates that Information C had the strongest appeal to a direct immune user of Information B and has the least attractiveness to an inactive user of the posted Information B.

**4.3 Influencing factors analytics: information release and dissemination**

In order to make a qualitative and quantitative analysis of the delay in transmission, we now introduce some additional indices, shown in Figure 6, and show how these can be used t]]]]]]]]]]der different effects of the posted information on newly posted information when the posted information has reached a quasi steady-state or is still in its outbreak period.

- *The outbreak peak $F_{2max}$*: the maximum of curve $F_2$, which reflects the peak user values of the newly posted information.
- *The final size $C_{2s}$*: the stable state of curve $C_2$, which gives the final size of the total number of users of the newly posted information.
- *the outbreak time $t_{2b}$, the end time $t_{2e}$ and the duration $t_{2i}$*: the definition depends on the outbreak threshold $F_2^*$ set in advance, so that $F_2(t_{2b}) = F_2^* = F_2(t_{2e})$. Here, $t_{2b}$ denotes the outbreak time of the newly posted information, $t_{2e}$ denotes the end time, and $t_{2i} = t_{2e} - t_{2b}$ denotes the duration of the newly posted information transmission. These time indices will help us judge the start and end of the newly posted information transmission.
- *The outbreak velocity $V_{2o}$ and the declining velocity $V_{2d}$*: the definition depends on $V_{2o} = (F_{2max} - F_2^*)/(t_{2max} - t_{2b})$ and $V_d = (F_{2max} - F_2^*)/(t_{2e} - t_{2max})$ when $F_2(t) = F_{2max}$ and $t_{2max}$ is definite, which reflects the speed of the outbreak and the decline of the newly posted information.

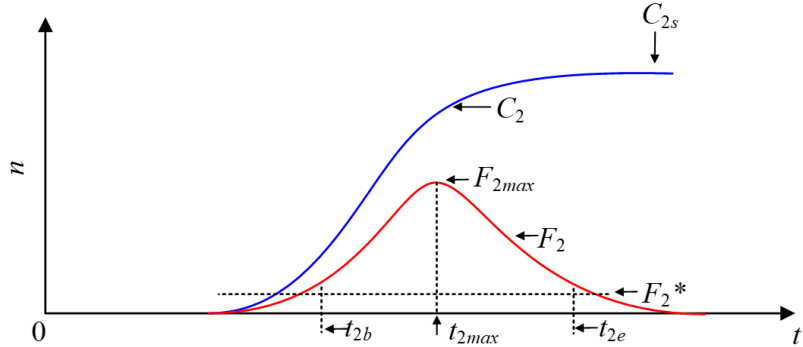

**Figure 6.** Some summative indices of a newly posted information that cross-propagating with an old information.

To further analyze the different parameters responsible for LTI DT-SFI, we performed an analysis of partial rank correlation coefficients (PRCCs) [30] to evaluate the sensitivity based on 1000 samples for various input parameters against the threshold condition. According to the histogram and scatter diagram of $R_0^1$ dependence, when the correlation is positive, it means that with the increase of the value of the parameter, the corresponding index value will increase; on the contrary, when the correlation is negative, the index will decrease as the parameter decreases. Figures 7-10 give the PRCC results and PRCC scatter plots with indices $\mathfrak{R}_o$, $F_{2max}$, $C_{2\infty}$, $t_{2b}$, $t_{2i}$, $t_{2max}$, $V_{2o}$, $V_{2d}$ with nine parameters ($\beta_{21}$, $\beta_{22}$, $\beta_{23}$, $p_2$, $\alpha_2$, $m_{21}$, $m_{22}$, $m_{23}$, $S_{20}$) of the newly posted information in LTI DT-SFI, respectively.

Figure 7 shows the effect of parameters on the public opinion reproduction ratio $\mathfrak{R}_o$ of the delay in transmission in LTI DT-SFI. $\mathfrak{R}_o$ is strongly positively affected by the average exposure rate $\beta_{22}$, the mild exposure attractiveness index $m_{22}$, the average exposure rate $\beta_{23}$, the un-exposure attractiveness index $m_{23}$ and the forwarding probability $p_2$, and strongly negatively affected by the average immune rate $\alpha_2$. The positive correlation effects of the parameters $\beta_{21}$ and $m_{21}$ are relatively weak. Overall, strategies to increase the parameters $\beta_{22}$, $\beta_{23}$, $p_2$, $m_{22}$, $m_{23}$ and initial value $S_{20}$ or to decrease the $\alpha_2$ can enhance the transmission capability of the newly posted information.

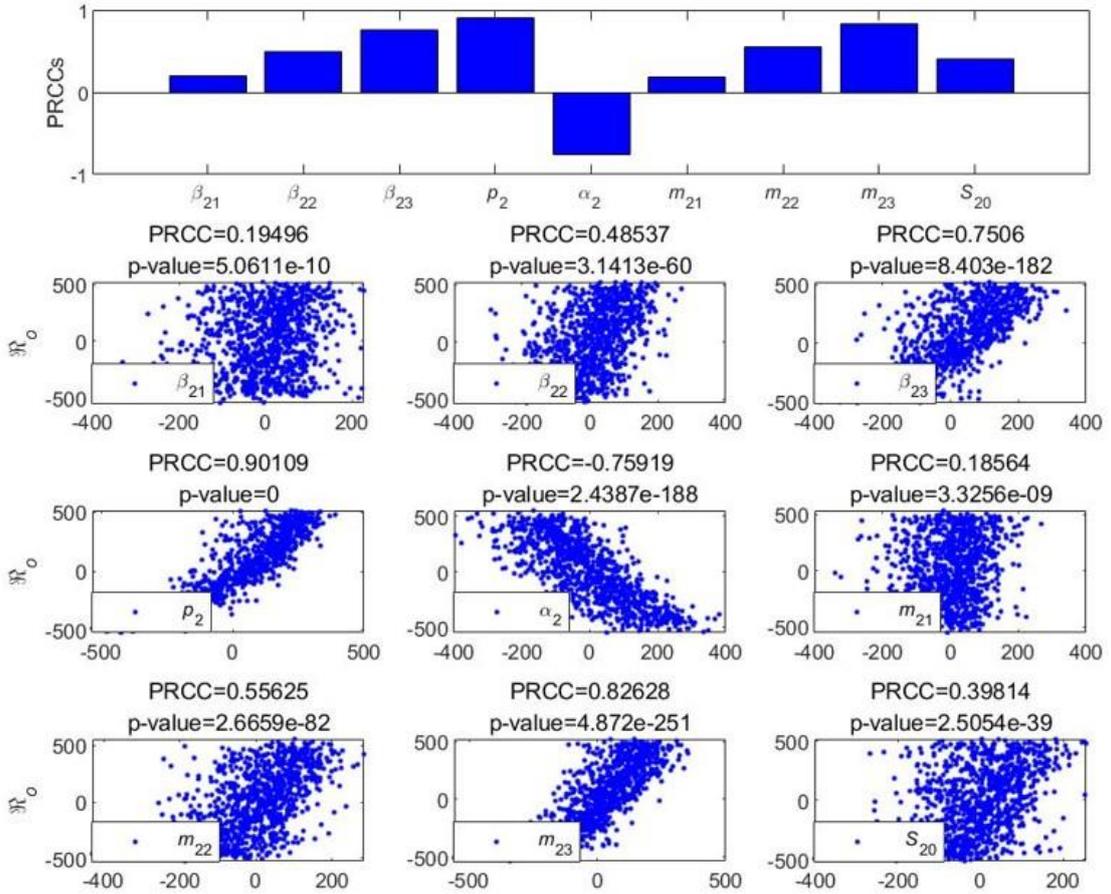

**Figure 7.** PRCC results and PRCC scatter plots with indices $\mathfrak{R}_o$ of different parameters of the newly posted information in LTI DT-SFI

From Figure 8, the parameters have a similar effect on the forwarding peak $F_{2max}$ and the cumulative forwarding population $C_{2\infty}$. The un-exposure attractiveness index $m_{23}$, the forwarding probability $p_2$ and the initial value $S_{20}$ of susceptible individuals have a decisive positive influence on the forwarding peak value $F_{2max}$ and the final size $C_{2\infty}$ of delayed information propagation. The effects of the extensive and mild exposure attractiveness parameters which portray the participation of the population of the posted information are very weak. The above results indicate that the time interval is long between the two delays in transmission information since the new information posted in a quasi-steady state into the propagation, at this time, most individuals who have been exposed to the posted information have entered the immune state. In addition, most individuals will no longer care about the relevant content due to the possibility of forgetting or leaving the social network platform. The above conclusions show that when the posted information enters the steady state, the effect of the individuals who have contacted the posted information is not obvious. Therefore, the information transmission can be promoted by influencing the number of new susceptible population.

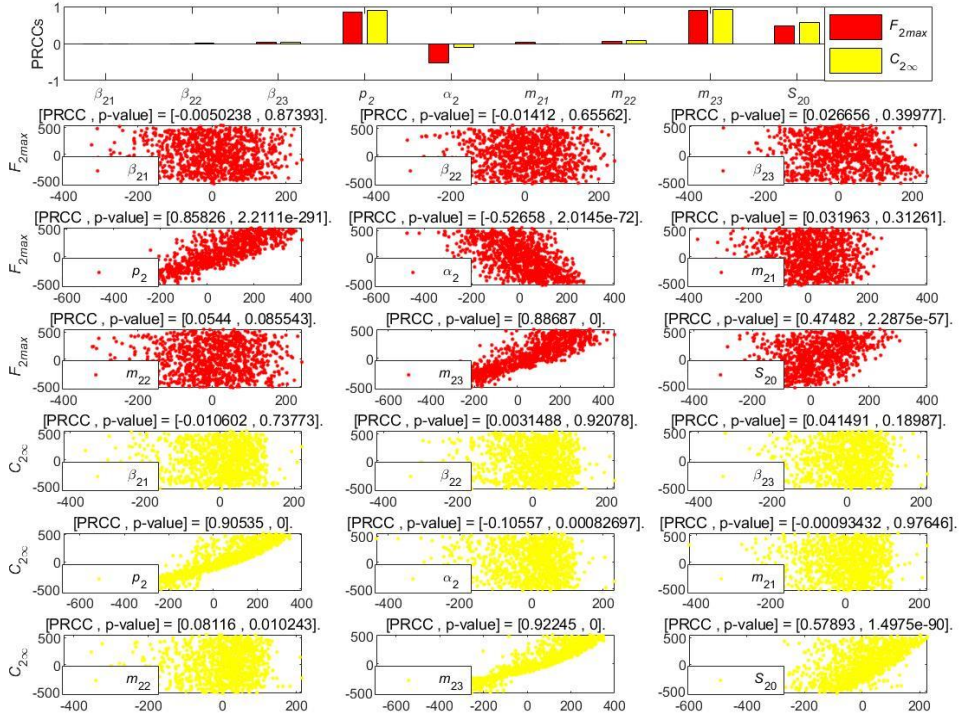

**Figure 8.** PRCC results and PRCC scatter plots with indices $F_{2max}$ and $C_{2\infty}$ of different parameters of the newly posted information in LTI DT-SFI

Figure 9 shows the effect of parameters on the climax time $t_{2max}$, the outbreak time $t_{2b}$ and the duration $t_{2i}$ of the delay in transmission. After mastering the influencing factors of $t_{2b}$ and $t_{2i}$, the end time of transmission $t_{2e}$ can be calculated. The climax time $t_{2max}$, the outbreak time $t_{2b}$ and the duration $t_{2i}$ are negatively affected by parameters $\beta_{23}$, $p_2$ and $m_{23}$ in the same way. In comparison, these parameters have the least impact on $t_{2b}$, especially $m_{23}$. The parameter $m_{22}$ has a weak negative correlation effect on each time index, and the parameter $\alpha_2$ is the main factor to control the duration $t_{2i}$ which plays a strong negative correlation effect.

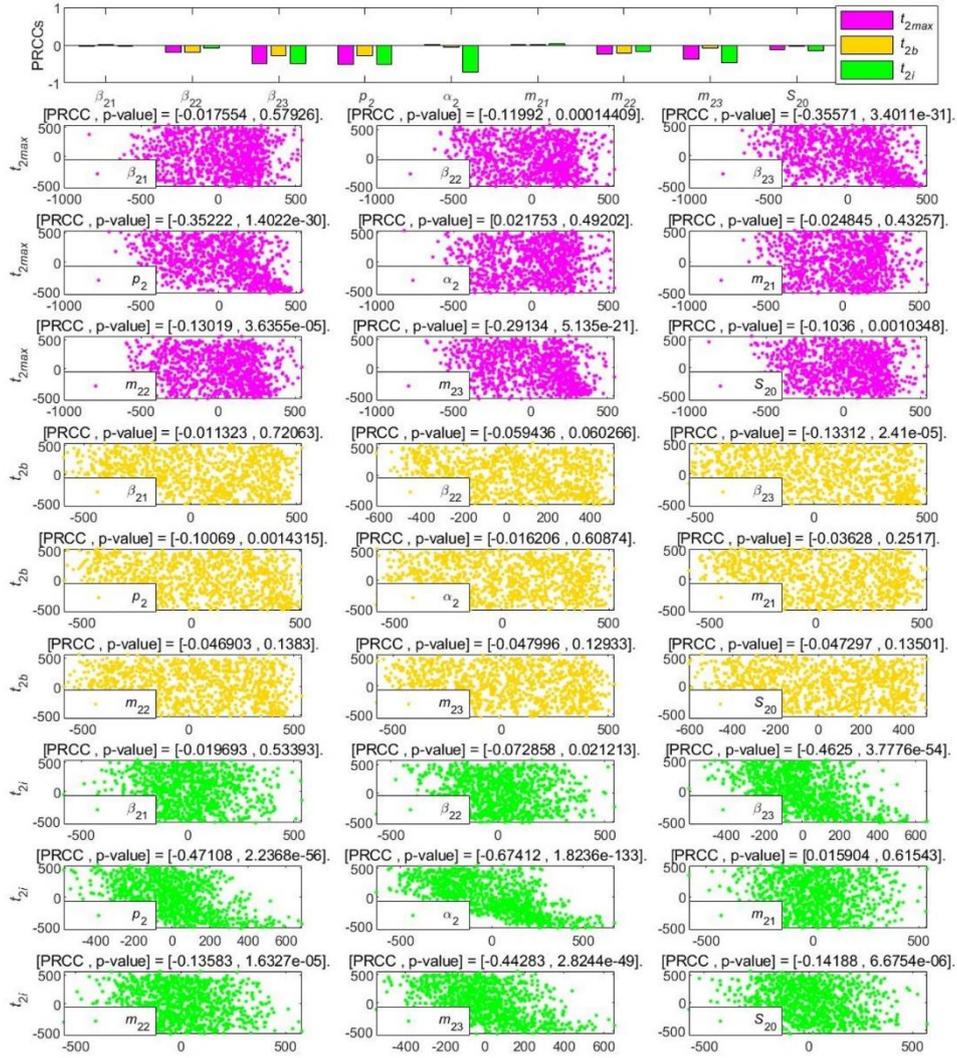

**Figure 9.** PRCC results and PRCC scatter plots with indices $t_{2max}$, $t_{2b}$ and $t_{2i}$ of different parameters of the newly posted information in LTI DT-SFI

From Figure 10, the un-exposure attractiveness index $m_{23}$ and the forwarding probability $p_2$ have major positive correlation effects on the outbreak velocity $V_{2o}$ and the declining velocity $V_{2d}$, and the initial value $S_{20}$ of susceptible individuals has a mild positive effect on these two indices. Moreover, the parameter $\alpha_2$ has a strong negative effect on the $V_{2o}$. In addition, the effects of other parameters on the velocities are not important. That is to say, the $V_{2o}$ and $V_{2d}$ will increase accordingly when the parameters $m_{23}$, $p_2$ and initial value $S_{20}$ increase. At the same time, the $V_{2o}$ increases with the reduction of parameter $\alpha_2$. By contrast, the effect of $m_{23}$ on the velocities is greater.

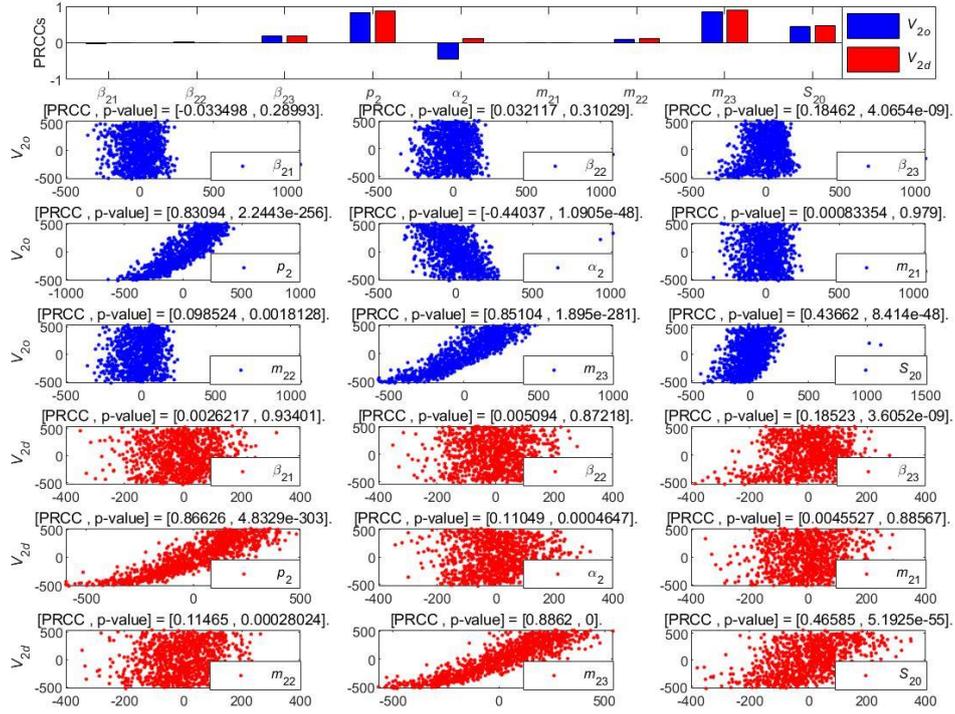

**Figure 10.** PRCC results and PRCC scatter plots with indices $V_{2o}$ and $V_{2d}$ of different parameters of the newly posted information in LTI DT-SFI

Our LTI DT-SFI model concentrates on the influence of the average exposure rates and attractiveness indexes on the instantaneous forwarding population $F_2(t)$ and the cumulative forwarding population $C_2(t)$ as shown in Figures 11-12 respectively, and the variation of parameters over time determines the propagation indices. By comparing and analyzing the influence of average contact rates and attractiveness indexes in Figure 11 and Figure 12 with the variation of one parameter while fixing other parameters, $\beta_{23}$ and $m_{23}$ have a similar overall trend of the effects on the instantaneous forwarding population $F_2(t)$ and the cumulative forwarding population $C_2(t)$ of the new information. With the increase of the parameters of $\beta_{23}$ and $m_{23}$, the outbreak will accelerate, and the instantaneous number of individuals in the forwarding state can reach a higher peak and the final size will be larger. In addition, the average exposure rate of $\beta_{22}$ and the mild exposure attractiveness index $m_{22}$ have a weak positive influence on the final size of the cumulative forwarding quantity, and have no obvious influence on the propagation times and velocities. In contrast, the average exposure rate $\beta_{21}$ and the extensive exposure attractiveness index $m_{21}$ had no significant effect on large interval delay in transmission based on forwarding. All the above key parameters have no significant effect on the outbreak time, climax time, and duration of the long-delayed cross-information transmission based on forwarding, which was also consistent with the results of PRCCs.

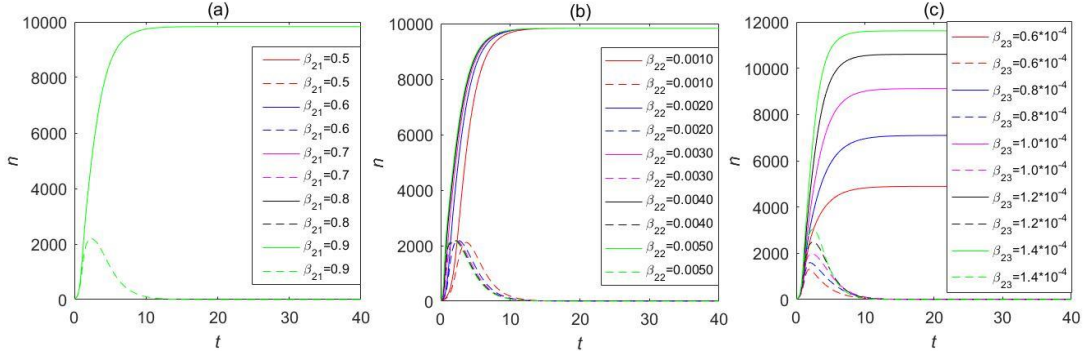

**Figure 11.** The influence of the average exposure rates on the instantaneous forwarding population $F_2(t)$ and the cumulative forwarding population $C_2(t)$ in LTI DT-SFI.

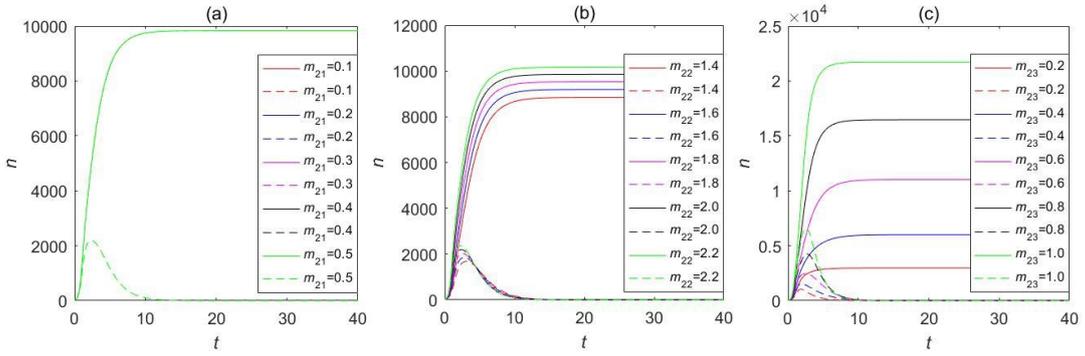

**Figure 12.** The influence of the attractiveness indexes on the instantaneous forwarding population $F_2(t)$ and the cumulative forwarding population $C_2(t)$ in LTI DT-SFI

## 5. Short delay in transmission susceptible-forwarding-immune dynamics model

### 5.1 Model description

Our comprehensive short interval delay in transmission susceptible-forwarding-immune (STI DT-SFI) dynamics model based on the forwarding quantity is shown in Figure 13. In this model, we include two phases: In phase 1, a stand-alone information (information 1) is spreading corresponding to phase 1 in LTI DT-SFI. In phase 2, a piece of newly posted information (information 2) is posted at $t_\tau$ during an outbreak period of the posted information. Here, we also divide the population into three groups: S-population ($S$), F-population ($F_1, F_2$) and I-population ($I_{1+}, I_{1-}, I_2$). In particular, we think of the susceptible state of both the posted information and the new information as a whole ($S$). $t_{in}$ is the post time of newly posted information.

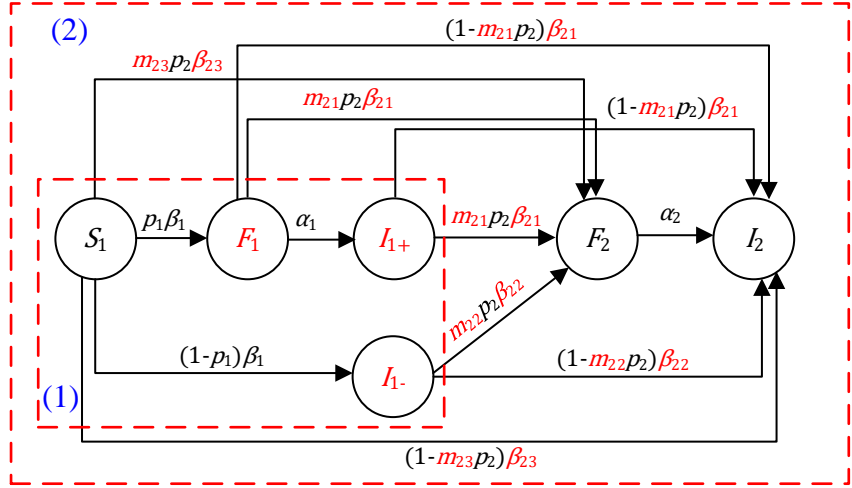

**Figure 13**. A schematic diagram to illustrate information spreading, when the post time of the newly posted information is during the outbreak period of the posted information.

**Phase 1. Post a stand-alone information**: The model is consistent with that at phase 1 of LTI DT-SFI developed in section **4.1.**

**Phase 2. Posting a new information during the outbreak period of the posted information:** Considering that the new information is posted during the outbreak period of the old information, we develop our phase 2 model to describe the concurrent dynamic process of the two related pieces of information. We consider the difference between the population in active forwarding state or the immune state out of the active period and the population in the direct immune state of the first (old) piece of information. Here, we set

- *an extensive exposure attractiveness index*, for the individuals in forwarding state who have forwarded the posted information but still in their active forwarding period and the individuals in immune state who have forwarded the old information but are no longer in their active forwarding period to indicate that this population will be attracted by the new information due to the relevance of the two pieces of information;
- *a mild exposure attractiveness index*, for the direct immune population to portray that the population will be attracted due to a moderate contact;
- an *un-exposure attractiveness index*, for the integrated susceptible population to describe that the population will be attracted when they have never read the related information.

Assuming that the number of users ($N_3$) who can contact the information in the process of information propagation on Sina-microblog remains unchanged, we introduce three states of the population of the newly posted information: the susceptible state ($S_1$) which includes the users who can be exposed to the old information and the new information, the forwarding state of the new information ($F_2$), the immune state ($I_2$). The parameters are shown in the following table 7.

**Table 7**. Parameters definition

| Parameter | Interpretation |
|---|---|
| **Attractiveness parameters, stratified by the exposure to the old information** | |
| $m_{21}$ | The extensive exposure attractiveness index that an inactive user of state $I_{1+}$ becomes a forwarding user of state $F_2$. |

| $m_{22}$ | The mild exposure attractiveness index that a direct immune user of state $I_{1-}$ becomes a forwarding user of state $F_2$. |
|---|---|
| $m_{23}$ | The un-exposure attractiveness index that a new susceptible user of state $S_2$ becomes a forwarding user of state $F_2$. |
| **Transmission parameters associated with the different attractiveness** | |
| $\beta_1$ | The average exposure rate that the susceptible users can contact the first information. |
| $\beta_{21}$ | The average exposure rate that the inactive users of the old information can contact the newly posted information. |
| $\beta_{22}$ | The average exposure rate that the direct immune users of the old information can contact the newly posted information. |
| $\beta_{23}$ | The average exposure rate that the new susceptible users can contact the newly posted information. |
| $p_1$ | The probability that the susceptible users will forward the first information. |
| $p_2$ | The probability that the exposed users will forward the newly posted information. |
| $\alpha_2$ | The average rate at which a user in the forwarding state of newly posted information becomes inactive to forwarding, where $1/\alpha_2$ is the average duration a forwarding user remains active in forwarding newly posted information. |

Each user may have a unique state, with $S_1(t)$, $F_1(t)$, $I_{1+}(t)$, $I_{1-}(t)$, $F_2(t)$ and $I_2(t)$ denoting the number of users in the susceptible, forwarding and immune state when $t > 0$, respectively. We obtain the following STI DT-SFI dynamics model in phase 2:

$$\begin{cases} dS_1(t)/dt = -\beta_1 S_1 F_1 - \beta_{23} S_1 F_2 \\ dF_1(t)/dt = p_1 \beta_1 S_1 F_1 - \beta_{21} F_1 F_2 - \alpha_1 F_1 \\ dI_{1+}(t)/dt = -\beta_{21} I_{1+} F_2 + \alpha_1 F_1 \\ dI_{1-}(t)/dt = (1-p_1)\beta_1 S_1 F_1 - \beta_{22} I_{1-} F_2 \\ dF_2(t)/dt = m_{21} p_2 \beta_{21} F_1 F_2 + m_{21} p_2 \beta_{21} I_{1+} F_2 + m_{22} p_2 \beta_{22} I_{1-} F_2 + m_{23} p_2 \beta_{23} S_1 F_2 - \alpha_2 F_2 \\ dI_2(t)/dt = (1-m_{21}p_2)\beta_{21} F_1 F_2 + (1-m_{21}p_2)\beta_{21} I_{1+} F_2 + (1-m_{22}p_2)\beta_{22} I_{1-} F_2 \\ \quad + (1-m_{23}p_2)\beta_{23} S_1 F_2 + \alpha_2 F_2 \end{cases} \quad (8)$$

The mass action in phase 2 can be interpreted as follows: An active forwarding user will contact an average number of $\beta_{21} N_3$ inactive immune users and forwarding users of posted information per unit time and the probability of a user is an inactive immune user and a forwarding user are $I_{1+}(t)/N_3$ and $F_1(t)/N_3$, so an active forwarding user will contact $\beta_{21} I_{1+}(t)$ inactive immune users and $\beta_{21} F_1(t)$ forwarding users, among which $m_{21} p_2 \beta_{21} I_{1+}(t) F_2(t)$ and $m_{21} p_2 \beta_{21} F_1(t) F_2(t)$ will choose to forward the new information however $(1-m_{21}p_2)\beta_2 I_{1+}(t) F_2(t)$ and $(1-m_{21}p_2)\beta_{21} F_1(t) F_2(t)$ will not; an active forwarding user will contact an average number of $\beta_{22} N_3$ direct immune users of posted information per unit time and the probability of a user is a direct immune user is $I_{1-}(t)/N_3$, so an active forwarding user will contact $\beta_{22} I_{1-}(t)$ direct immune users, among which $m_{22} p_2 \beta_{22} I_{1-}(t) F_2(t)$ will choose to forward the new information and $(1-m_{22}p_2)\beta_{22} I_{1-}(t) F_2(t)$ will not; an active forwarding user will contact an average number of $\beta_{23} N_3$ susceptible users and the probability of a user is a susceptible user is $S_1(t)/N_3$, so an active forwarding user will contact $\beta_{23} S_1(t)$ susceptible users, among which $m_{23} p_2 \beta_{23} S_1(t) F_2(t)$ will choose to forward the new information and $(1-m_{23}p_2)\beta_{23} S_1(t) F_2(t)$ will not.

The forwarding quantity of the newly posted information is given by

$$C_2(t) = \int_0^t (m_{21}p_2\beta_{21}F_1F_2 + m_{21}p_2\beta_{21}I_{1+}F_2 + m_{22}p_2\beta_{22}I_{1-}F_2 + m_{23}p_2\beta_{23}S_1F_2)\,dt. \tag{9}$$

**The public opinion reproduction ratio** $\mathfrak{R}_o$: Considering the initial condition in phase 2, we can obtain the following public opinion reproduction ratio $\mathfrak{R}_o$. The new information entered at time $\tau$ during an outbreak period of the posted information, and $F_2'(0) = (m_{21}p_2\beta_{21}F_{1\tau} + m_{21}p_2\beta_{21}I_{1+,\tau} + m_{22}p_2\beta_{22}I_{1-,\tau} + m_{23}p_2\beta_{23}S_{10\tau} - \alpha_2)F_2(0)$. The population will never take off if

$$F_2'(0) = (m_{21}p_2\beta_{21}F_{1\tau} + m_{21}p_2\beta_{21}I_{1+,\tau} + m_{22}p_2\beta_{22}I_{1-,\tau} + m_{23}p_2\beta_{23}S_{10\tau} - \alpha_2)F_2(0) < 0$$

due to the decrease of $S_{10\tau}$. It is therefore natural to introduce

$$\mathfrak{R}_o = \frac{m_{21}p_2\beta_{21}F_{1\tau} + m_{21}p_2\beta_{21}I_{1+,\tau} + m_{22}p_2\beta_{22}I_{1-,\tau} + m_{23}p_2\beta_{23}S_{10\tau}}{\alpha_2} \tag{10}$$

as the STI DT-SFI reproduction ratio.

In the same way, the $\mathfrak{R}_o$ of STI DT-SFI denotes the comprehensive public opinion generated by the newly posted Weibo starting at the outbreak period of the posted information. When the reproduction ratio $< 1$, it means that the new public opinion will decline. The reproduction ratio $> 1$ indicates that the new public opinion will grow exponentially initially.

### 5.2 Data fitting

**Parameter estimation**:

To use our model to explore some distinctions of the qualitative behaviors for prediction, we use the LS method to estimate the STI DT-SFI model parameters and the initial data of our model. The vector is set as $\Theta_3 = (p_1, \beta_1, \alpha_1, p_2, \beta_{21}, \beta_{22}, \beta_{23}, m_{21}, m_{22}, m_{23}, \alpha_2, S_{10})$, and the corresponding numerical calculation based on the parameter vectors for $C_1(t)$ and $C_2(t)$ are denoted by $f_{C_1}(k, \Theta_3)$ and $f_{C_2}(k, \Theta_3)$, respectively. The LS error function

$$LS = \sum_{k=0}^T |f_{C_1}(k, \Theta_3) - C_{1k}|^2 + \sum_{k=0}^T |f_{C_2}(k, \Theta_3) - C_{2k}|^2 \tag{11}$$

is used in our calculation, where $C_{1k}$ and $C_{2k}$ denote the actual cumulative forwarding populations of the posted information and the newly posted information, here, $n = 1,2,3$ represents the different phases, $k = 0,1,2,...$ is the sampling time. $n = 1,2,3$. We estimate the parameters of our **STI DT-SFI** model with the data of Information A and Information B.

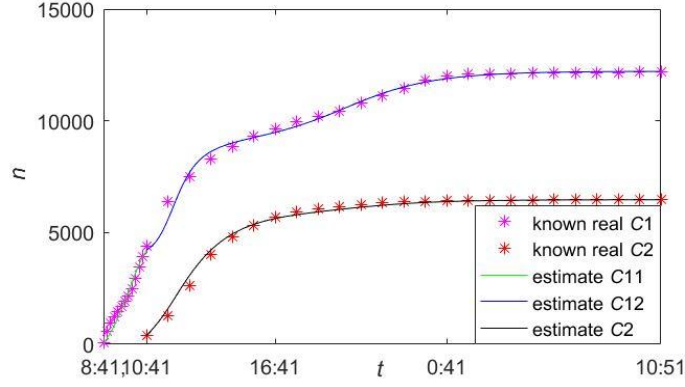

**Figure 14.** The data fitting results of Information A and Information B

In the data fitting of **STI DT-SFI**, we use the same method as that in Section **4.2** to fit the data of Information A and Information B. As shown in Figure 14, we performed data fitting of Information A and Information B on the real data in Table 1-2, where the pink star denotes the actual cumulative number of forwarding users of Information A, the red star denotes the actual cumulative number of forwarding users of Information B, the green line and the blue line denotes the estimated cumulative number of forwarding users in the early and later period of Information A, respectively, and the black line denotes the estimated cumulative number of forwarding users of Information B. It can be seen that our STI DT-SFI model achieves accurate estimation.

**Table 8**. Values of some important parameters, estimated for Information A

| name  | $p_1$  | $\beta_1$              | $\alpha_1$ | $S_{10}$              |
|-------|--------|------------------------|------------|-----------------------|
| value | 0.9823 | $8.2700\times 10^{-5}$ | 3.9986     | $5.1682\times 10^{4}$ |

**Table 9**. Values of important parameters, estimated for Information B

| Name  | $p_1$  | $\beta_1$              | $\alpha_1$ | $p_2$  | $\beta_{21}$ | $\beta_{22}$ | $\beta_{23}$           | $m_{21}$ | $m_{22}$ | $m_{23}$ | $\alpha_2$ | $S_0$                 |
|-------|--------|------------------------|------------|--------|--------------|--------------|------------------------|----------|----------|----------|------------|-----------------------|
| Value | 0.0091 | $3.6601\times 10^{-4}$ | 3.4777     | 0.0788 | 0.0037       | 0.8184       | $6.8834\times 10^{-5}$ | 0.0406   | 0.0109   | 0.1868   | 1.9159     | $7.4439\times 10^{6}$ |

Table 8 gives some important values of parameter (relevant to the early period of the outbreak) estimation of Information A, and Table 9 gives some important values of parameter estimation for the later period data of Information A and all data of Information B. We can see in phase 2, when Information B was posted during the outbreak period of Information A, the average exposure rate $\beta_{21}$ and $\beta_{22}$ are much larger than $\beta_1$ and $\beta_{23}$, which indicates that users who have been exposed to Information A will contact Information B at a greater rate than new susceptible users. In addition, the un-exposure attractiveness index $m_{23}$ is the largest among the three attractiveness indexes, since the time interval between two information posted is small, and people who have not been exposed to relevant information may have a greater interest in new information, the outbreak of Information B has the strongest appeal to susceptible users.

**5.3 Influencing factors analytics: information release and dissemination**

To further analyze the impact of different parameters in STI DT-SFI model for the cross-propagation dynamics, we performed PRCCs to analyze the relationship between the influence and the range of variation of parameters on the indexes. Figures 15-18 give the PRCC results and PRCC scatter plots with indices $\Re_o$, $F_{2max}$, $C_{2\infty}$, $t_{2b}$, $t_{2i}$, $t_{2max}$, $V_{2o}$, $V_{2d}$ mentioned above with nine parameters ($\beta_{21}$, $\beta_{22}$, $\beta_{23}$, $p_2$, $\alpha_2$, $m_{21}$, $m_{22}$, $m_{23}$, $S_{10}$) of the newly posted information in STI DT-SFI, respectively.

The average exposure rate $\beta_{23}$, the un-exposure attractiveness index $m_{23}$, the forwarding probability $p_2$ and the initial value $S_{10}$ make strong contributions to the public opinion reproduction ratio $\Re_o$ positively, and the average immune rate $\alpha_2$ has a strong negative effect on it as shown in Figure 14. The positive correlation effects of parameters $\beta_{21}$, $\beta_{22}$, $m_{21}$ and $m_{22}$ are relatively weak. In general, strategies that can affect the parameters $\beta_{23}$, $m_{23}$, $p_2$ and the initial value $S_{10}$ to increase or the parameter $\alpha_2$ to decrease can increase the initial propagation capacity of the newly posted information. On the other hand, we can decrease the parameters $\beta_{23}$, $m_{23}$, $p_2$ and the initial value $S_{10}$ to reduce the initial propagation ability of the new information.

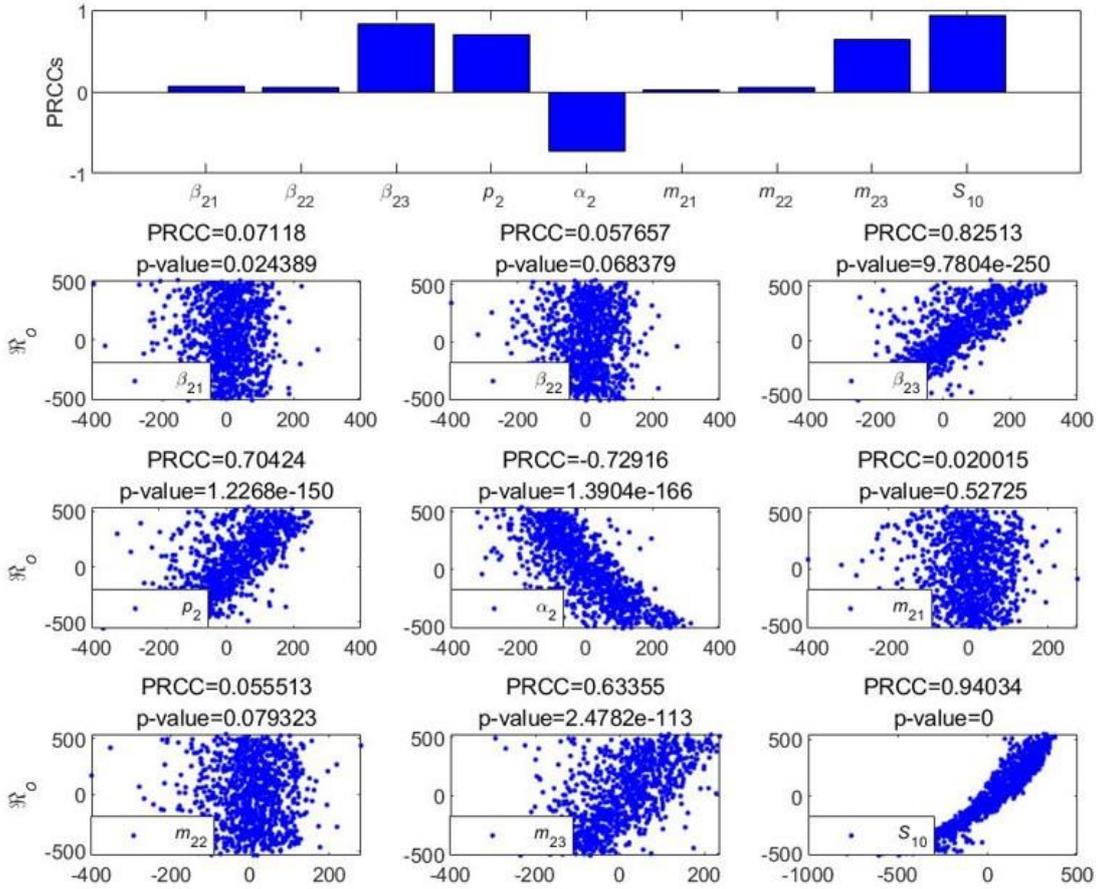

**Figure 15.** PRCC results and PRCC scatter plots with indices $\Re_o$ of different parameters of the newly posted information in STI DT-SFI.

The average contact rate of $\beta_{21}$, $\beta_{23}$, the forwarding probability $p_2$, the mild exposure attractiveness index $m_{22}$, the un-exposure attractiveness index $m_{23}$ and the initial value $S_{10}$ of susceptible individuals have strong positive impacts on the high peak $F_{2max}$ and the final size $C_{2\infty}$ as shown in Figure 16. In contrast, $S_{10}$ plays a major role, and the impact of $\beta_{22}$ and $m_{21}$ are less significant. The above results show that individuals who have been exposed to but have not forwarded the posted information are more sensitive to the new information with mild exposure attractiveness due to the understanding of the former information. In addition, the average immune rate $\alpha_2$ has a strong negative effect on the $F_{2max}$.

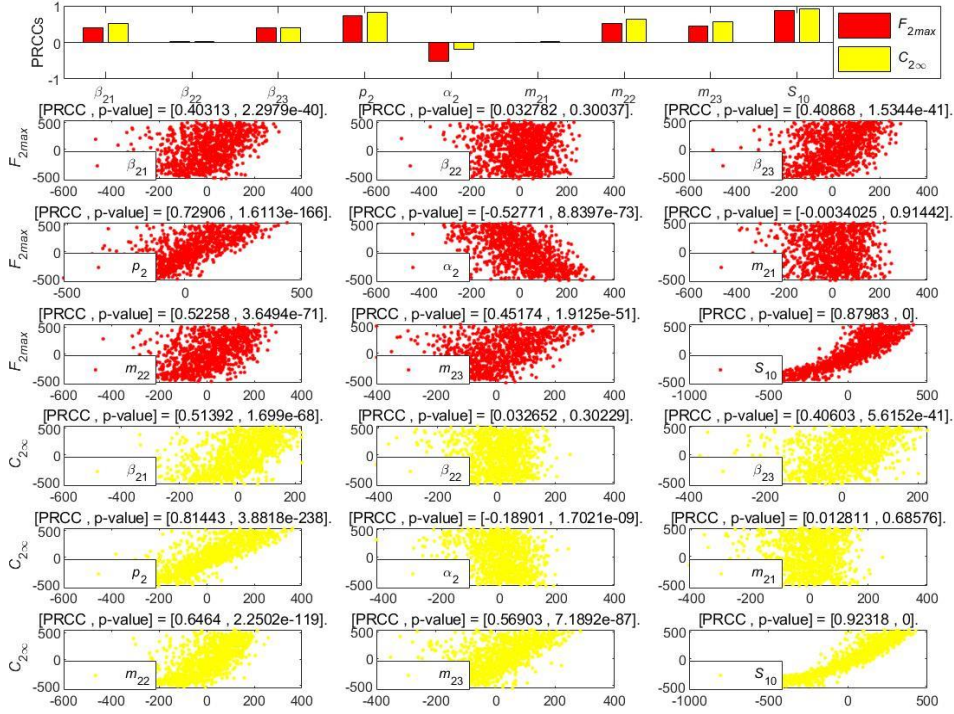

**Figure 16.** PRCC results and PRCC scatter plots with indices $F_{2max}$ and $C_{2\infty}$ of different parameters of the newly posted information in STI DT-SFI.

Figure 17 indicates that the influence of each parameter on $t_{2max}$ and $t_{2b}$ are not obvious, and the parameters $\beta_{23}$, $\alpha_2$ and the initial value $S_{10}$ have negative effects on the duration $t_{2i}$, and the $\beta_{21}$ has a weak positive effect on it. This means that the average contact rate at which users in the susceptible state can contact the second information is the most important factor affecting the duration $t_{2i}$ of delay in transmission. The smaller the average contact rate is, the longer the duration of new information transmission will be within a certain range, and then slowing down the development of information transmission.

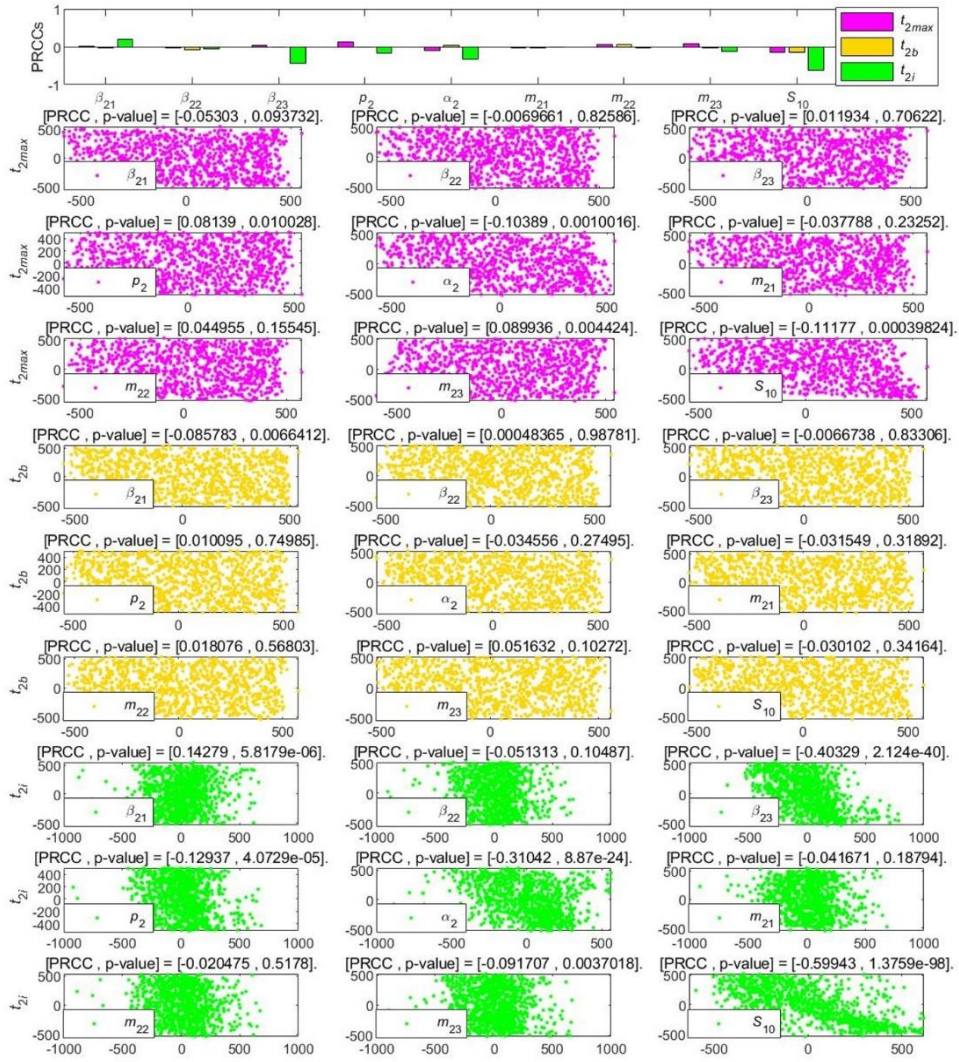

**Figure 17.** PRCC results and PRCC scatter plots with indices $t_{2max}$, $t_{2b}$ and $t_{2i}$ of different parameters of the newly posted information in STI DT-SFI.

Figure 18 shows the PRCCs results of the outbreak velocity $V_{2o}$ and the declining velocity $V_{2d}$ of STI DT-SFI based on forwarding under multi-parameters changes. From the results, the average exposure rate $\beta_{21}$, $\beta_{23}$, the forwarding probability $p_2$, the mild exposure attractiveness index $m_{22}$, un-exposure attraction index $m_{23}$, and the initial value $S_{10}$ of susceptible individuals make strong positive contributions on $V_{2o}$ and $V_{2d}$. The average exposure rate $\beta_{22}$ and the extensive exposure attractiveness index $m_{21}$ has no significant effect on the velocities. That is to say, the outbreak velocity $V_{2o}$ and the declining velocity $V_{2d}$ can be increased with the increase of the parameters $\beta_{21}$, $\beta_{23}$, $p_2$, $m_{22}$, $m_{23}$ and the initial value $S_{10}$. On the contrary, if the parameters decrease, the propagation velocities will slow down.

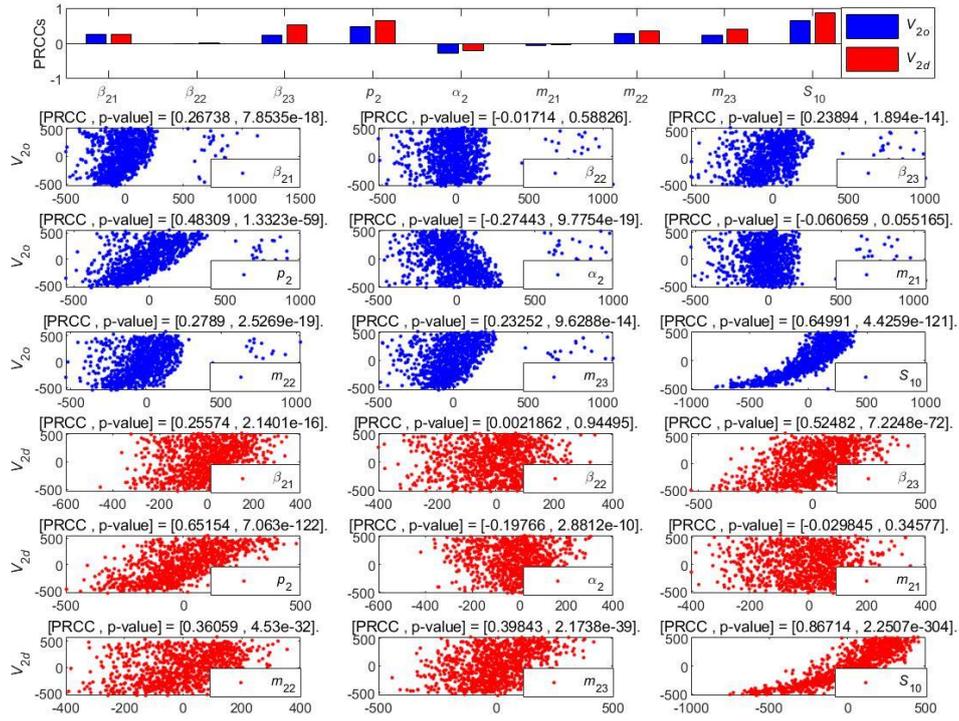

**Figure 18.** PRCC results and PRCC scatter plots with indices $V_{2o}$ and $V_{2d}$ of different parameters of the newly posted information in STI DT-SFI.

Here, we also take into consideration the influence of the average exposure rate and attractiveness parameters on the instantaneous forwarding population $F_2(t)$ and the cumulative forwarding population $C_2(t)$ of STI DT-SFI as shown in Figures 19-20 respectively. The comparative analysis shows that the larger the average contact rate and attractiveness indexes are, the larger the instantaneous forwarding quantity and the cumulative forwarding quantity are. The final size is also affected. the average exposure rate of $\beta_{21}$, $\beta_{23}$, the mild exposure attractiveness index $m_{22}$, and the un-exposure attractiveness index $m_{23}$ are the main influencing factors of STI DT-SFI model, and they can play a significant role in the final size of the newly posted information within a certain range. So priority must be placed on controlling these parameters. In addition, the extensive exposure attractiveness index $m_{21}$ has only a small magnitude of effects, while the effect of parameter $\beta_{22}$ is significant and has a relatively obvious impact. The impact of each parameter on the outbreak timing and increasing/declining velocities is negligible, which is consistent with the results of PRCCs.

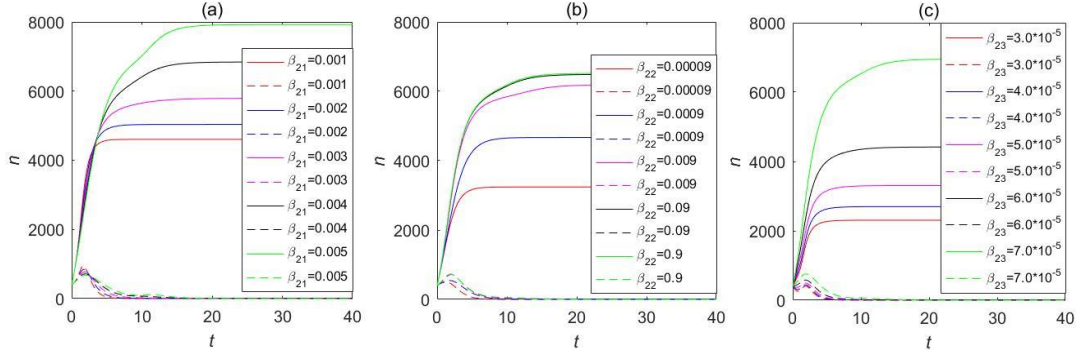

**Figure 19.** The influence of the average exposure rates on the instantaneous forwarding population $F_2(t)$ and the cumulative forwarding population $C_2(t)$ in STI DT-SFI.

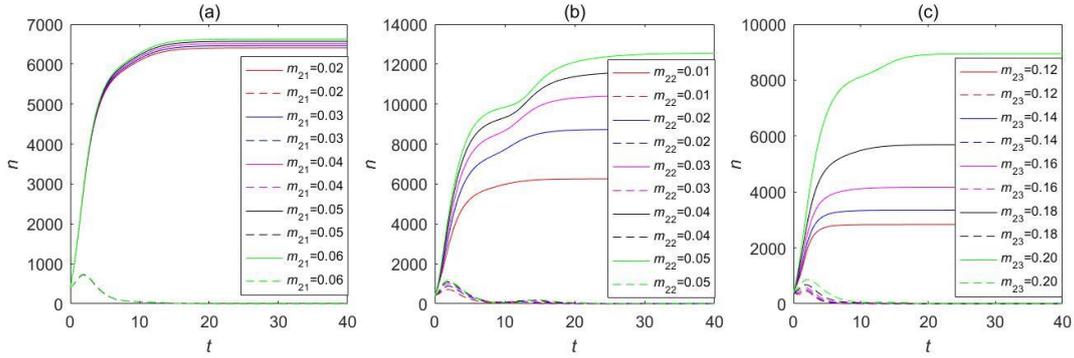

**Figure 20.** The influence of the attractiveness parameters on the instantaneous forwarding population $F_2(t)$ and the cumulative forwarding population $C_2(t)$ in STI DT-SFI.

## 6. Comparison and Discussion

Figure 21 shows the trend of the cumulative forwarding users of Information B and Information C when the newly posted information is posted during the steady-state period of the posted information in Tables 2-3. The time lag with which the newly posted information is posted has a significant impact on the process of public opinion dissemination and the final size of the cumulative number of forwarding users. If the newly posted information is posted during the quasi steady state period of opinion propagation, then the earlier the newly posted information is posted, the earlier the cumulative forwarding users will peak, though the final size of the cumulative forwarding users will be close to each other. This, in conjunction with our parameter sensitivity analysis results, shows that for the LTI DT-SFI situation, the un-exposure attractiveness index and the average exposure rate $\beta_{23}$ are the key elements to promote the cross-propagation and that once reaching the quasi steady state, the timing of posting the new information has an insignificant impact on the final size of forwarding users.

In contrast, Figure 22 shows the trend of the cumulative forwarding users of Information A and Information B when the newly posted information is posted during the outbreak period of the posted information in Table 1-2. The time lag with which the newly posted information is posted has noticeable impact on both the dynamic process of public opinion dissemination and the final size of the cumulative number of forwarding users. If the newly posted information can be posted during the outbreak period of the old information, then the earlier the new information is posted, the greater the cumulative forwarding population and the final size of the cumulative forwarding users will be.

This, combined with our parameter sensitivity analysis results, shows that in the STI DT-SFI case, the mild exposure attractiveness index $m_{22}$, the average exposure rates $\beta_{21}$ and $\beta_{23}$, and the un-exposure attractiveness index $m_{23}$ can all directly influence the interaction between information posted sequentially to increase the "heat" (popularity) of the newly posted information.

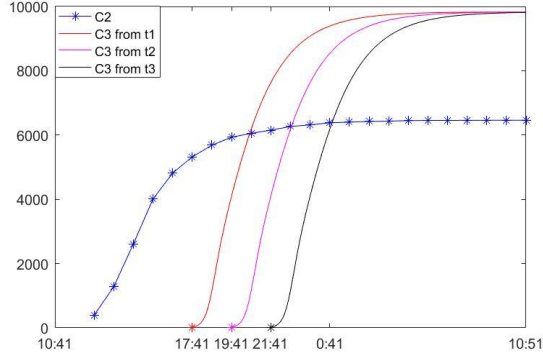

**Figure 21** An illustration of a public opinion dissemination process with newly posted information posted with different time lags but during the quasi steady state period of the posted information.

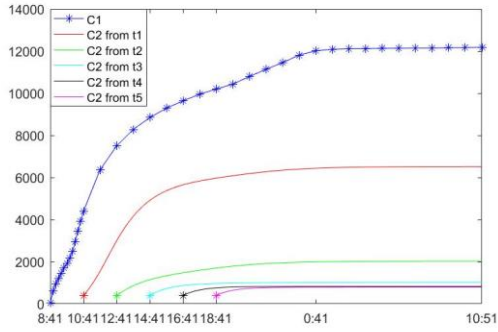

**Figure 22** An illustration of a public opinion dissemination process with newly posted information posted at different times: the newly posted information is posted during the outbreak period of the posted information.

Our model-based analysis recommends strategies on how different parameters should be adjusted to achieve the best information dissemination outcomes. For two pieces of relevant information separated by a relative long posting lag, strategies to increase the average exposure rate $\beta_{23}$ and the un-exposure attractiveness index $m_{23}$ are recommendations. These strategies can be achieved if opinion leaders with a large number of followers can participate in the information co-propagation. On the contrary, reducing the public's attention to a new piece of information can be achieved by efforts in delaying the posting of the new information, and/or by effectively reducing the potential correlation between the two pieces of information (reducing values of the correlation parameters $\beta_{21}$, $\beta_{22}$, $m_{21}$, $m_{22}$). Additionally, if our goal is for the final size of the cumulative forwarding users of the new information not be impacted by the relevant information already posted online, the new information should be posted during the quasi steady state period of the posted information.

For two pieces of information with a short interval between postings, we recommend to develop strategies to alter the interaction between the information for effectively managing the information transmission indices we introduced. If we aim to make the new information outbreak faster with a large peak value of forwarding, we should increase the average exposure rate $\beta_{21}$ and mild exposure attractiveness index $m_{22}$ by persuading the original post owner to post or forward the information earlier during the outbreak period of the posted information when the posted information has obtained certain public attention, and increase the relevance and attraction of the newly posted information to the forwarding users or immune users of the posted information. Alternatively, we should persuade some opinion leaders to forward the new information along with their insights to $\beta_{23}$ and $m_{23}$.

## 7. Conclusions

It is often that relevant information is posted sequentially in fast evolving public health events such as the COVID-19 pandemic. Consequently, modeling the impact of delay in cross-transmission or co-propagation is significant to identify the best strategies to communicate key public messages through social media. In this study, we proposed and examined two classes of models, large delay in transmission susceptible-forwarding-immune (LTI DT-SFI) dynamics model and the short delay in transmission susceptible-forwarding-immune (STI DT-SFI) dynamics model based on the forwarding users in Weibos, and we parametrized our models using real data related to COVID-19 pandemic in the Chinese Sina-Microblog. Our goal is to use these parametrized models to understand the influence of different time lags in the information posting on the co-propagation of related information in the Microblog.

Our model formulation focused on the transmission mechanism of information in the social network, where a new Weibo may be posted in different phases—outbreak phase and/or quasi steady state phase—of some relevant Weibo already posted. Our goal is to examine the impact of post timing, in relation to the old information, of the new information on its peak value and final size of forwarding users. As we have shown, this impact depends on the correlation of the old and new information, and on the phase of the old information transmission when the new information is posted. It is hoped that our DT-SFI dynamics model fills in some theoretical gap about optimizing information posting strategies to maximize communication efforts to deliver key public health messages to the public for better outcomes of public health emergency management.

## Acknowledgments

The work was supported by the National Natural Science Foundation of China (Grant numbers: 61801440), the Natural Science and Engineering Research Council of Canada, the Canada Research Chair Program (JWu), the Fundamental Research Funds for the Central Universities and the High-quality and Cutting-edge Disciplines Construction Project for Universities in Beijing (Internet Information, Communication University of China), State Key Laboratory of Media Convergence and Communication, Communication University of China.